\documentclass[preprint,aps,12pt,preprintnumbers,nofootinbib,superscriptaddress] {revtex4}

\usepackage{graphicx}
\usepackage{amssymb}
\usepackage{amsmath}
\usepackage{epstopdf}
\usepackage{subfigure}

\def\a{\alpha}
\def\ad{{\dot \alpha}}
\def\b{\beta}
\def\bd{{\dot \beta}}
\def\c{\gamma}
\def\cd{{\dot \gamma}}

\def\t{\tilde}

\newcommand\ket[1]{{| {#1} \rangle}}
\newcommand\sqet[1]{{| {#1} ]}}
\newcommand\bra[1]{{\langle {#1} | }}
\newcommand\sqra[1]{{[ {#1} | }}

\newcommand{\nl}{\nonumber\\}

\begin{document}

\title{Spinor Helicity and Dual Conformal Symmetry in Ten Dimensions}

\author{Simon Caron-Huot}
\affiliation{Institute for Advanced Study, Einstein Drive, Princeton, NJ}
\author{Donal O'Connell}
\affiliation{Institute for Advanced Study, Einstein Drive, Princeton, NJ}
\affiliation{The Niels Bohr International Academy,
            The Niels Bohr Institute,
            Blegdamsvej 17,
            DK-2100 Copenhagen,
            Denmark}

\date{\today}

\begin{abstract}

The spinor helicity formalism in four dimensions has become a very useful tool both for understanding the structure of amplitudes and also for practical numerical computation of amplitudes. Recently, there has been some discussion of an extension of this formalism to higher dimensions. We describe a particular implementation of the spinor-helicity method
in ten dimensions. Using this tool, we study the tree-level S-matrix of ten dimensional super Yang-Mills theory,
and prove that the theory enjoys a dual conformal symmetry. Implications
for four-dimensional computations are discussed.

\end{abstract}

\maketitle

\section{Introduction}

There has been considerable progress in the study of amplitudes in
quantum field theories in recent years. One technique which has allowed
for a lot of progress in four dimensions has been the spinor-helicity
method. This method was developed independently by three different
groups~\cite{XuZhangChang,GunionKunszt,KleissStirling}, building on
a critical insight of~\cite{De Causmaecker:1981by}, namely, that the
simplicity of helicity amplitudes containing explicit polarisation vectors
built from spinors explained the remarkable cancellations discovered in
single bremsstrahlung processes~\cite{Berends:1981rb}. This technology
allows one to write extremely simple, gauge invariant expressions for
$n$ point amplitudes in four dimensions, of which the most celebrated
example is surely the Parke-Taylor amplitudes~\cite{ParkeTaylor}.
It also blends beautifully with modern techniques to compute tree
amplitudes \cite{BerendsGiele,Cachazo:2004kj,Britto:2004ap,Britto:2005fq}.

In recent years, the spinor-helicity method has been extended to higher
dimensions. This was first achieved in six dimensions~\cite{Cheung:2009dc}
and later a similar (but not quite identical) formalism was constructed in
arbitrary dimensions~\cite{Boels:2009bv}. The six-dimensional formalism
was supersymmetrized~\cite{Dennen:2009vk} and has been recently used
in~\cite{Bern:2010qa, Brandhuber:2010mm} to study the structure of loop
diagrams in six dimensions.

In the present paper, we will extend the method of~\cite{Cheung:2009dc} to ten dimensions. Thus, we will describe how spinor variables in ten dimensions which have explicit Lorentz and little group indices may be used to describe the momenta and polarisations of particles in ten dimensions. In addition, we will describe one method for
understanding on-shell supersymmetry in ten dimensions. We remark that the methods we describe in this paper allow one to entend the spinor-helicity formalism to any dimension.

Having constructed these general tools, we turn our attention to one very particular theory: $\mathcal{N} = 1$ super Yang-Mills theory in ten dimensions. This theory is, of course, closely related to $\mathcal{N} = 4$ super
Yang-Mills in four dimensions. More precisely, (color ordered) tree amplitudes in ten dimensions are in one to one correspondence
with scattering amplitudes in $\mathcal{N}=4$ on the Coulomb branch. The momenta along the 6 extra dimensions
dictate the pattern of Higgs vacuum expectation values in four dimensions.
The $\mathcal{N} = 4$ theory enjoys a fascinating symmetry which has been named dual conformal symmetry. This
symmetry of the on-shell scattering amplitudes emerged and was developed in a series of papers by several authors~\cite{Drummond:2006rz, Alday:2007hr,Bern:2006ew, Bern:2007ct, Drummond:2007aua, Brandhuber:2007yx, Drummond:2007cf, Drummond:2007au, Drummond:2008vq, drummondhennplefka, Berkovits:2008ic, brandhuber,arkanihamedetal}.
Working at tree level, we will demonstrate that the tree amplitudes of the ten dimensional theory also enjoy a dual conformal symmetry. We achieve this by defining the generator of special dual-conformal transformations, and showing that the symmetry is preserved by BCFW-type recursion relations.
This symmetry generator reduces to the dual conformal generator defined in
\cite{drummondhennplefka}
upon dimensional reduction. In addition, our proof of the dual conformal invariance of the ten dimensional tree amplitudes will closely follow the method used by~\cite{brandhuber} in four dimensions.

While we only work at tree level, our results have two broad
implications for loop computations in lower dimensions.\footnote{Or
course, $\mathcal{N} = 1$ super Yang-Mills theory in ten dimensions
is anomalous.} Firstly, to the extent that tree amplitudes determine
the integrand of loop diagrams, the dual conformal symmetry of the
theory in ten dimensions implies that the loop integrands in all lower
dimensions enjoy an appropriate dual conformal invariance. Indeed, while
this work was being completed, a paper appeared~\cite{Bern:2010qa}
noting the vanishing of integral coefficients in six dimensions
which would violate this symmetry; the authors of~\cite{Bern:2010qa}
were then motivated to investigate dual conformal invariance in six
dimensions and even wrote down a generator of special dual-conformal
transformations in . The second implication of this work
is that the four dimensional theory can indeed be regulated by moving
onto the Coulomb branch of the theory~\cite{Alday:2009zm}, in such a
way as to preserve covariance under the dual conformal transformations.
From our perspective, the regulator masses of the various particles can
be thought of as components of momenta in the extra dimensions. Thus, our
dual conformal generators act on these masses, as in~\cite{Alday:2009zm}.

The structure of this paper is as follows. In section~\ref{sec:spinors} we describe the spinor variables which will be our basic tools. We will construct expressions for momenta and polarisation vectors before examining the on-shell supersymmetry algebra. Next, section~\ref{sec:BCFW} deals with the topic of BCFW recursion~\cite{Britto:2004ap,Britto:2005fq} in ten dimensions. In particular, our interest will be in how to deform the bosonic and fermionic variables describing our particles in order to ensure that the BCFW recursion relations can be used. We turn our attention exclusively to $\mathcal{N}=1$ super Yang-Mills theory in ten dimensions in section~\ref{sec:symmetries}. In this section, we demonstrate that the theory has an SO(2,10) dual conformal symmetry. Finally, in section~\ref{sec:concl} we discuss our results and conclude. Our appendices contain some more technical details.

\section{Spinor Variables}
\label{sec:spinors}

We work in a mostly negative metric convention. In ten dimensions it is possible to choose a basis of the Clifford algebra so that
$\sigma^\mu$ and $\tilde \sigma^\mu$ are symmetric matrices, obeying
\begin{equation}
 \sigma^\mu_{ab} \tilde \sigma^{\nu bc} + \sigma^\nu_{ab} \tilde\sigma^{\mu bc} = 2\delta_a^c \eta^{\mu\nu}.
\end{equation}
The spinor indices $a=1\ldots 16$.
The goal of this section is to provide spinorial expressions for momenta and polarisation vectors. A key concept throughout will be the interaction of the ten dimensional Lorentz group and the SO(8) little group\footnote{More precisely, the subgroup of the SO(1,9) Lorentz group which fixes a lightlike vector $p$ is the group ISO(8) of Euclidean motions (rotations and translations) in eight dimensional space. As is usual~\cite{Weinberg:1995mt}, we assume that the generators of translations in this space annihilate physical states, so that only the SO(8) subgroup has a non-trivial action. It is this subgroup that we call the little group;
the non-compact generators in ISO(8) have trivial action on all the objects we consider and can be omitted.}. In fact the little group plays a more prominent role so we will now discuss eight dimensional spinors in some detail.

\subsection{The little group}

SO(8) has three distinct 8-dimensional representations: one vector,
one chiral and one antichiral spinor. All of these will play a role
in our story. In each case there is an inner product which we can
characterize by a symmetric tensor. Throughout the text of this
paper we will take this inner product to be the Kronecker symbol,
which is clearly a convenient choice while working in ten dimensions.
(We remark that this choice, proves less convenient when one wishes to dimensionally
reduce.)
Thus, the inner products will be $\delta_{\a \b}$
and $\delta_{\dot\a\dot\b}$. Once a basis is fixed, there is a
invariant tensor $\rho_{m\alpha\dot\alpha}$,
which generates the Clifford algebra
\begin{equation}
\rho_m \rho^T_n + \rho_n \rho^T_m = 2 \delta_{mn}.
\label{lgPauli}
\end{equation}
SO(8) is actually quite a special group because of triality symmetry which
interchanges its various 8-dimensional representations.
Let us anticipate that one
of the SO(8) representation will describe gauge bosons, another the
fermion states, and the third the supersymmetry generators which hop
between the two.

\subsection{Momenta}

Now we move on to consider on shell momenta. First, a word on notation: the spinors we consider will be charged under the Lorentz group as well as a little group. There are two kinds of spinor (chiral and antichiral) and, in addition, we will have to consider pairs of these spinors for each particle. Thus there is a danger of being overwhelmed with indices. We will use two notations for the spinors, and swap between them for convenience. That is, we shall denote a spinor for particle $i$ with momentum $p_i$ as either $\lambda_{i\alpha}^a$ or $\ket{i, \alpha}^a$ where $a = 1, \ldots, 16$ is a Lorentz spinor index, while $\alpha = 1, \ldots, 8$ is a little group spinor index. For antichiral spinors we shall use the notation $\tilde \lambda_{i \dot \alpha a}$ or $\sqet{i, {\dot \alpha}}_a$. We will frequently suppress the Lorentz spinor indices. With this information in hand, let us move on to see how the relevant spinors are constructed.

Given a momentum $p$ with $p \cdot p = 0$ we can find a basis of 8 spinors by solving the Weyl equations
\begin{equation}
\label{defSpinors}
p \cdot \sigma \ket{p, \a}= 0; \quad p \cdot \tilde \sigma \sqet {p, \ad}  = 0.
\end{equation}
We wish to construct a spinorial expression for the momentum. Consider the object $T_{\alpha \beta}^\mu = \bra{p, \alpha} \sigma^\mu \ket{p, \beta} = T_{\beta \alpha}^\mu$. One can show that this object is proportional to $p^\mu$, times a SO(8) invariant tensor.

Let us elaborate on why this is so, because we will use a similar argument
several times in what follows. The way this works is that the SO(8) little
group should be thought of as a subgroup of the SO(1,9) Lorentz group.
Indeed it can be embedded as the subgroup leaving invariant $p^\mu$ plus
some other arbitrary reference vector; all such embeddings are isomorphic
to each other. Thus one can decompose the ten dimensional vector into
irreducible SO(8) representations, and compare this decomposition with
the irreducible representations found in the symmetric tensor product of
two SO(8) spinors. In the present case, one finds that the only overlap
is the singlet representation. Thus, the only non-vanishing components
of $T_{\alpha\beta}^\mu$ are singlets in $\alpha\beta$.

Thus, it is possible to take the spinors to be orthogonal in a canonical sense.
To fix the normalisation convention, we choose to impose
\begin{equation}
p \cdot \tilde \sigma = \sum_\alpha \ket{p, \alpha} \bra{p, \alpha}. \label{completeness}
\end{equation}
It follows that $\bra{p, \alpha} \sigma^\mu \ket{p, \beta} = 2 \delta_{\alpha \beta} p^\mu$. 

We will also need a basis of antichiral spinors, which obey
\begin{equation}
 \sqra{p, \ad} \tilde \sigma^\mu \sqet{p, \bd} = 2 \delta_{\ad \bd} p^\mu.
\end{equation}
It is useful to note that all spinor contractions vanish, $s_{\a\ad}=\lambda_\alpha^a\tilde\lambda_{\dot\alpha a} =0$.
Thus the bases of chiral and antichiral spinors are completely independent.
The vanishing is because the product is a Lorentz scalar, and thus must be a little-group scalar by the above argument. But there is no room for a little-group scalar with this index structure. Thus $s_{\a \ad}$ vanishes. 

It may help to point out an important difference between eight and ten dimensional spinors, and their four and six dimensional counterparts. In four dimensions, the tensor product of two arbitrary spinors (one chiral, one antichiral) determines a null momentum. The counting of degrees of freedom is simple: there are two degrees of freedom in each spinor, minus one degree of freedom from the rescaling $\lambda \rightarrow z \lambda, \,\, \tilde \lambda \rightarrow \tilde \lambda / z$ which leaves the tensor product unchanged. The result is precisely the three degrees of freedom of the null momentum. In six dimensions, one simply takes the antisymmetric part of the two four component spinors; this combination is left invariant by an $SL(2, \mathbb{C})$ transformation of the two spinors, leaving the five degrees of freedom of six dimensional massless spinor. 

However, in eight and ten dimensions the situation is a little more subtle. In these cases, the tensor product of two spinors contain a wealth of $n$ forms. If we were to choose 8 completely random spinors, the completeness relation (\ref{completeness}) would
be spoiled by the appearance of a self-dual 5-form component on the right-hand side. This means that
the spinors $\lambda_i^\alpha$ obey nonlinear constraints.
As far as we know, the best way to solve these constraints is to begin by specifying $p^\mu$, and then find the $\lambda_i^\alpha$
as orthogonal solutions to the Weyl equation.

\subsection{Polarisation vectors}

%The following spinor identity will be useful in this construction.
%\begin{equation}
%\sigma_\mu {}_{a (b} \sigma^\mu{}_{cd)} = 0 = \tilde \sigma^{\mu a (b} \tilde \sigma_\mu{}^{cd)} .
%\end{equation}  .... Not even here!

Consider the object $\rho_{m \alpha \dot \alpha} \, \lambda_\alpha (\sigma^\mu \tilde \sigma^\nu - \sigma^\nu \tilde \sigma^\mu) \tilde \lambda_{\dot \alpha}$. This object has the property that it transforms as a vector under little group transformations, while it transforms as a two-form under Lorentz boosts. These are the same transformation properties as the field strength tensor $F_m^{\mu \nu}$ of a vector particle. It follows that
\begin{equation}
F^{\mu \nu}_m = \theta \rho_{m \alpha \dot \alpha} \, \lambda_\alpha (\sigma^\mu \tilde \sigma^\nu - \sigma^\nu \tilde \sigma^\mu) \tilde \lambda_{\dot \alpha}
\label{fieldStrength}
\end{equation}
where $\theta$ is a suitable normalisation constant. 

This observation allows us to deduce a convenient expression for the polarisation vectors, $\epsilon_m^\mu$,
of a particle, such that $F^{\mu\nu}_m = \epsilon_m^\mu p^\nu - \epsilon_m^\nu p^\mu$.
Let us choose some vector $n$ with the property that $p \cdot n \neq 0$. We can therefore arrange for the polarisation vectors to satisfy $n \cdot \epsilon_m = 0$. In this gauge, we find
\begin{equation}
\epsilon^\mu_m = \rho_{m\a\ad} \frac{\lambda_\alpha \, \sigma^\mu \tilde n \!\!\! / \, \tilde \lambda_{\dot \alpha}}{16 \, p \cdot n}.
\end{equation}
It is instructive to compute the field strength corresponding to $\epsilon^\mu_m$ explicitly:
\begin{align}
\epsilon_m^\mu p^\nu - \epsilon_m^\nu p^\mu &= \frac{1}{16 p \cdot n} \rho_{m\a\ad} \lambda_\a \left(p^\nu \sigma^\mu \tilde n \!\!\! /  - p^\mu \sigma^\nu \tilde n \!\!\! / \right)  \tilde \lambda_\ad \nonumber \\
&= \frac{1}{16 p \cdot n} \rho_{m\a\ad} \lambda_\a \left(\frac{1}{2} \sigma^\nu \tilde p \!\!\! / \sigma^\mu \tilde n \!\!\! /  - p^\mu \sigma^\nu \tilde n \!\!\! / \right)  \tilde \lambda_\ad \nonumber \\
&= \frac{1}{16 p \cdot n} \rho_{m\a\ad} \lambda_\a \left(-\frac{1}{2} \sigma^\nu \tilde \sigma^\mu p \!\!\! / \tilde n \!\!\! / \right)  \tilde \lambda_\ad \nonumber \\
&= \frac{1}{16} \rho_{m\a\ad} \lambda_\a \left( \sigma^\mu \tilde \sigma^\nu \right)  \tilde \lambda_\ad \nonumber  \\
&= F_m^{\mu \nu}.
\end{align}
Thus, we learn that $\theta = 1/32$.  This computation also shows
that $\epsilon^\mu_m$ depends on $n$ only modulo gauge transformation,
so that the physical state corresponding to $\epsilon^\mu_m$ does not
depend on $n$.

We have chosen to normalise the vectors so that
\begin{equation}
\epsilon_m \cdot \epsilon_n = - \delta_{mn}.
\end{equation}
It is also straightforward to check that the polarisation vectors are a complete set in the sense that
\begin{equation}
\sum_m \epsilon^\mu_m \epsilon_m^\nu = - \left( \eta^{\mu\nu} - \frac{p^\mu n^\nu + p^\nu n^\mu}{p \cdot n} 
 + \frac{p^\mu p^\nu n^2}{(p\cdot n)^2} \right).
\end{equation}

It is possible to use a little group spinor notation for all the above objects.
Consider the object $\lambda_\alpha (\sigma^\mu \tilde \sigma^\nu - \sigma^\nu \tilde \sigma^\mu) \tilde \lambda_{\dot \alpha}$.  This transforms as a two-form under Lorentz boosts, while for its little group transformation properties,
on purely group theoretical grounds one might expect
that it contains a little group vector and a three form.  However, a spacetime two form cannot give rise to a little group three form.
Therefore the three form component is absent. This means that we can write $F_{\a\ad}^{\mu\nu}$ interchangeably for
$F^{\mu\nu}_m$:
\begin{equation}
F_m^{\mu\nu} = \frac18 \rho_{m \alpha \dot \alpha} \, F_{\alpha \dot \alpha}^{\mu\nu},  \quad \quad\quad
F_{\alpha \dot \alpha}^{\mu\nu} = F_m^{\mu\nu} \, \rho_{m \alpha \dot \alpha}.
\end{equation}
Similarly, the corresponding polarisation vector contains no little group three form, and we can write
\begin{equation}
\epsilon_m^\mu = \frac18 \rho_{m \alpha \dot \alpha} \, \epsilon_{\alpha \dot \alpha}^\mu,  \quad \quad\quad
\epsilon_{\alpha \dot \alpha}^\mu = \epsilon_m^\mu \, \rho_{m \alpha \dot \alpha}
\end{equation}
where
\begin{equation}
\epsilon^\mu_{\a\ad} = \frac{\lambda_\alpha \, \sigma^\mu \tilde n \!\!\! / \, \tilde \lambda_{\dot \alpha}}{2 \, p \cdot n}.
\end{equation}

Of course, vector particles in ten dimensions are not the only particles of interest. Appropriate polarisation tensors for massless spin $\frac12$ particles are provided by the spinors $\tilde \lambda_\ad$ for antichiral fermions and $\lambda_\a$ for chiral fermions. In our discussion of supersymmetry below, we will choose the spinor partners of the Yang-Mills vectors to be antichiral spinors. Then their polarisations will be
\begin{equation}
 \psi_{a \ad} = \tilde\lambda_{a \ad}.
\end{equation}
One can also construct polarisation tensors for higher spin objects. For
example, in ten dimensions one can consider a particle which is a
self-dual four-form of the little group and has a self-dual five-form
fieldstrength in ten dimensions. Such a particle has a gauge-dependent
polarisation tensor $\gamma^{mnop}_{\a \b} \lambda_\a \sigma^{\mu \nu
\rho \sigma} n \!\!\! / \lambda_\b$ and gauge independent field strength
$\gamma^{mnop}_{\a \b} \lambda_\a \sigma^{\mu \nu \rho \sigma \tau}
\lambda_\b$.

\subsection{Supersymmetry}

How does supersymmetry act on on-shell single particle states?  The boson
and fermion states live in the vector and spinor representation of the
little group SO(8), respectively.  Supersymmetry must intertwin these
two representations.  The only possibility consistent with Lorentz
covariance is
\begin{equation}
 Q^a = \lambda^a_{\a} \Gamma_{\a}.   \label{defSUSY}
\end{equation}
Let us explain the notation.  The SO(8) Gamma matrices carry three indices,
one in each of the 8-dimensional representation of SO(8). We can choose
\begin{equation}
\Gamma_\alpha = \begin{pmatrix}
0 & \rho_\alpha \\
\rho_\alpha^T & 0 
\end{pmatrix}.
\end{equation}
The matrices $\rho$ are the SO(8) Clifford algebra matrices introduced in Eq.~\eqref{lgPauli}. In particular, these matrices have indices in each of the three 8 dimensional representations of SO(8).
The action of $\Gamma$ sends the boson state $\ket{m}$ to $\Gamma_{\a\dot\b}^m \ket{\dot\b}$
and the fermion state $\ket{\dot\b}$ to $\Gamma_{\a\dot\b}^m \ket{m}$.
Using the Clifford algebra one checks that
indeed $ \{ Q^a,Q^b\} = 2 p \cdot \tilde \sigma^{ab}$.

It is instructive to verify explicitly the consistency of (\ref{defSUSY})
with the above polarisation choices. The complete multiplet of
polarisations is $(\tilde \lambda_{a\dot \alpha},  \epsilon_m^\mu)$. Let
us begin by examining the action of the supersymmetry generator on the
fermion polarisation $\tilde \lambda$. Using the definition of the
supercharge, and projecting onto the $m$th bosonic state, we find
\begin{equation}
(Q^a \, \tilde \lambda_{b})_m = \lambda^a_\alpha \, \rho_{m \alpha \dot \alpha} \, \tilde \lambda_{b \dot \alpha}.
\end{equation}
Now, let us examine the object $\lambda^a_\alpha \, \rho_{m \alpha \dot \alpha} \, \tilde \lambda_{b \dot \alpha}$.
Notice that this object is a little group vector. In the Clebsch-Gordon decomposition of $\lambda^a \tilde \lambda_b$ in terms of $n$ forms, the only object which can transform as a little group vector is the spacetime
2 form. Normalising correctly, we find
\begin{equation}
 \lambda^a_\alpha \, \rho_{m \alpha \dot \alpha} \, \tilde \lambda_{b \dot \alpha} = -\frac{1}{4} F^{\mu \nu}_m (\sigma_{\mu}\tilde \sigma_{\nu}- \sigma_{\mu}\tilde \sigma_{\nu}){}_b{}^a,
\end{equation}
so that the supersymmetry variation of the fermion is the fieldstrength of the vector, as expected. 

Next, let us examine the action of the supercharge on the vector polarisation. We compute
\begin{align}
(Q \epsilon^\mu)_{\ad} &= \lambda_{\alpha} \rho_{m \alpha \dot \alpha} \epsilon^\mu_m \nonumber \\
&=  \lambda_\alpha \epsilon^\mu_{\alpha \dot \alpha} \nonumber \\
&= \frac{\lambda_\alpha \lambda_\alpha \, \sigma^\mu \tilde n \!\!\! / \, \tilde \lambda_{\dot \alpha}}{2 p \cdot n} \nonumber \\
&= \frac{ \tilde p \!\!\! / \, \sigma^\mu \tilde n \!\!\! / \tilde \lambda_{\dot \alpha}}{2 p \cdot n} \nonumber \\
&= - \tilde \sigma^\mu \tilde \lambda_{\dot \alpha} + \frac{p^\mu}{p \cdot n} n \!\!\! / \tilde \lambda_{\dot \alpha} .
\end{align}
Thus, the susy variation of the vector yields the spinor, modulo a gauge transformation.

Conventional four-dimensional on-shell superspace fits in the present language as follows.
The little group is SO(2) 
% I removed this because I have no idea what it means
%and there is one Gamma matrix for each of the two handed-ness.
%Furthermore, 
and massless particles form doublets of states.
On such a doublet the Gamma matrices can be represented as
\begin{equation}
 \Gamma_i^+ = \left( \begin{array}{ll}   0 & \eta_i \\ 0 & 0\end{array}
 \right),
\quad
 \Gamma_i^- = \left( \begin{array}{ll}   0 & 0 \\ \frac{\partial}{\partial \eta_i} & 0\end{array}
 \right),
 \label{analogy}
\end{equation}
where $\eta_i$ is an anticommuting Grassmann variable.  The S-matrix can then be viewed as a
function of the $\eta_i$. With $\mathcal{N}$ four-dimensional supersymmetries,
a supermultiplet is described by $\mathcal{N}$ $\eta_i$ variables.

An on-shell superspace formalism for six dimensions was described in
\cite{Dennen:2009vk}.  It was found to be necessary to explicitly break
the SU(2)$\times$SU(2) R-symmetry in order to have the correct number of
fermion variables.  We believe that a similar formalism will also work
in eight dimensions, where the required four Grassman variables can be
taken to transform under the 4 of the little group SO(6)$\cong$SU(4);
these variables will be charged under the U(1) R-symmetry which will
not be manifest.

In ten dimensions, however, it does not appear possible to introduce $\eta$-variables while maintaining manifest
the full Lorentz symmetry of the theory.  Indeed, to account for the $16=2^4$ states one would need 4
$\eta$-variables, but the smallest representations of the little group gives 8 variables.
It thus appears that the best superspace one could achieve in ten dimensions would depend on the same data as
a dimensional reduction to 8 dimensions.
For this reason, we shall not use $\eta$-variables in ten dimensions but will stick to the Gamma matrix notation,
which preserves all the symmetries of the theory.

\section{BCFW}
\label{sec:BCFW}

Since our goal is to understand the symmetry structure of the S-matrix,
we need a recipe for computing the S-matrix elements. We will construct
the (tree) S-matrix by the BCFW method.  Thus, we need to understand
how to deform the variables specifying a state to allow for the on-shell
recursion.

\subsection{Bosonic deformation}

It is useful to recall the BCFW deformation for massless particles in four dimensions
\begin{equation}
 \hat\lambda_1 = \lambda_1 - z\lambda_n, \quad
 \hat {\tilde \lambda}_n = \tilde \lambda_n + z\tilde \lambda_1.  \label{4dshift}
\end{equation}
Notice that $z$ implicitly has a little group charge, which is U(1) in four dimensions.
In higher dimensions, we choose to consider a variable $z$ which has no little group transformation property. The little group transformations are handled by choosing a matrix $M_{\b \a}$.
Then the deformation is
\begin{align}
\hat \lambda_{1\a} &= \lambda_{1\a} -z \lambda_{n\b}  M_{\b\a}, \\
\hat \lambda_{n\b} &= \lambda_{n\b} +z M_{\b\a}\lambda_{1\a},
\end{align}
where $z$ is a complex variable.
The matrix $M_{\b\a}$ has some constraints.
To have a linear deformation of the momenta we impose that $M_{\b\a}M_{\c\a}=0=M_{\b\a}M_{\b\c}$.
The other constraint is that we want to preserve the orthogonality relations between the spinors $\lambda$.  This requires
\begin{equation}
\lambda_{n\c} M_{\c \a} \sigma^\mu \lambda_{1\b} +\lambda_{n\c} M_{\c \b} \sigma^\mu \lambda_{1\a}
= 2\delta_{\a \b} q^\mu,
 \label{orthoQ}
\end{equation}
where the vectors are shifted as
$
 \hat p_1 = p_1 - z q$ and $\hat p_n = p_n + z q.
$

The vector $q^\mu$ is by construction null and orthogonal to $p_n$ and $p_1$.  In fact, $M$ and $q$ are completely equivalent
since one can show
\begin{equation}
\label{eqforM}
  M_{\b\a} = - \frac{\lambda_{n\b} q \!\!\! / \, \lambda_{1\a}}{2p_n\cdot p_1}.  
\end{equation}
Corresponding to these deformations of the chiral spinors, there is a deformation of the antichiral spinors given by
\begin{align}
\hat{\tilde \lambda}_{1\ad} &= \tilde \lambda_{1\ad} -z \t \lambda_{n\bd}  \t M_{\bd\ad}, \\
\hat{\t \lambda}_{n\bd} &=\t \lambda_{n\bd} +z \t M_{\bd\ad}\t \lambda_{1\ad},
\end{align}
where the matrix $\t M_{\bd \ad}$ is given by
\begin{equation}
\t M_{\bd \ad} = - \frac{\tilde \lambda_{n\bd} \tilde q\!\!\!/\, \tilde \lambda_{1\ad}}{2 p_n \cdot p_1}.
\end{equation}

Thus the data of the deformation is a null vector in the 8-dimensional subspace orthogonal
to $p_1$ and $p_n$.  It is illuminating to relate this vector to other vectors that we already know in that subspace.  Namely,
the polarisation vectors of particle 1 in the gauge where $p_n\cdot\epsilon_1=0$.  In fact, in this gauge, one finds that
\begin{equation}
q^\mu = V_{m} \epsilon_{1 m}^\mu   \label{qfromV}
\end{equation}
where $V_{m} = \frac18 \rho_{m \a \ad} (\tilde \lambda_{1 \ad} \cdot \lambda_{n \b} ) M_{\b \a}$.
The constraint that $q$ be null is the same as the constraint that $V$ be null.
Thus, the choice of a deformation vector is precisely given by the choice of a null SO(8) vector $V_m$.
Another useful observation is that the object $V_{\a\ad}=(\tilde \lambda_{1 \ad} \cdot \lambda_{n \b} ) M_{\b \a}$,
which a priori could contain a vector and a 3-form with respect to the SO(8) of particle 1, as above, contains only
the 1-form $V_m$.  

In a non-supersymmetric theory, one must correlate one of the polarisation
vectors with the deformation in order to ensure that the deformed amplitude
falls off at large $z$. However, since we will focus on a supersymmetric theory,
this issue will not be important for us.

\subsection{Supersymmetric deformation}

In a four-dimensional supersymmetric theory, the shift Eq. \eqref{4dshift} can be supplemented with the
fermion shift $\hat{\eta}_n=\eta_n + z\eta_1$.
The defining property of this shift is that it ensures that the shift commutes with supersymmetry:
\begin{eqnarray}
  \hat Q  - Q =
  \left[(\lambda_1-z\lambda_n)\eta_1 + \lambda_n(\eta_n+z\eta_1)\right]
 - \left[\lambda_1\eta_1 + \lambda_n\eta_n\right] = 0.
\end{eqnarray}
A similar, six-dimensional version of this shift has been given in \cite{Dennen:2009vk};
we are interested here in a ten-dimensional generalization.
Since we are not using a superspace formulation, it is useful first to rewrite the above in operator form
as $e^{z\eta_1\frac{\partial}{\partial \eta_n}}$.  Then, using the analogy
(\ref{analogy}), it is easy to find the corresponding shift in ten dimensions:
\begin{equation}
  A_z(p_1,p_2,\ldots,p_n) = e^{-\frac12 z \Gamma_{n\a} M_{\a\b} \Gamma_{1\b}}
        A(\hat p_1, p_2, \ldots, \hat p_n).
\end{equation}
It is easy to verify that $A_z$ is annihilated by $Q$ if and only if $A$ is:
this shift indeed commutes with supersymmetry. 

Thanks to supersymmetry, to check that the deformed super-amplitude vanishes at infinity
it suffices to consider just one specific polarisation
choice for particles 1 and $n$.  This is because using
supersymmetry we can always fix the polarisations of particles 1 and $n$ to be what we want.
Furthermore, exploiting Lorentz invariance, we can assume that particles 1 and $n$ and the deformation
vector lie in a four-dimensional subspace.  Then our deformation is exactly the same as the usual four
dimensional BCFW deformation. Now, the individual amplitudes which are components of the super-amplitude have
the same $z$ scaling in any dimension as in four dimensions~\cite{ArkaniHamed:2008yf}. Therefore, it is irrelevant
that some of the momenta in our amplitudes do not lie in the privileged four dimensions containing the deformation and momenta of particles 1 and $n$. Thus, the same proof that the super-amplitude falls off at
large $z$ in four dimensions~\cite{ArkaniHamed:2008gz,brandhuber} carries over to the present case.
Like in four dimensions, the supermultiplet inherits the scaling properties of its the best-behaved components,
which are not affected by the fermionic shifts.

Since the deformed amplitude vanishes at infinity, the original amplitude can be constructed from
its poles.  This establishes the following recursion relation \cite{Britto:2004ap,Britto:2005fq} (for color-ordered partial amplitudes):
\begin{equation}
  A_n = \sum_{k=3}^n e^{-\frac{z_0}2 \Gamma_{n\a} M_{\a\b} \Gamma_{1\b}}
    A_L(\hat p_1, \ldots, p_k,\hat p_L)
  \frac{1}{p_L^2}
    A_R(\hat p_R, p_{k+1},\ldots, \hat p_{n})  \label{treeBCFW}
\end{equation}
where $p_L=-p_R=-\sum_{j = 1}^k p_j$ and
the hatted bosonic variables, in the $k$th term, are shifted by an amount corresponding to
$z_0= p_{1 \ldots k}^2/2q{\cdot}p_{1 \ldots k} $.
This expresses the $n$-point amplitude in terms of products of strictly lower-point amplitudes.
The vanishing at infinity of the amplitude ensures that $A$ is independent of the
choice of deformed particles and $M_{\alpha\beta}$.

\begin{figure}[t]
\centering
\includegraphics{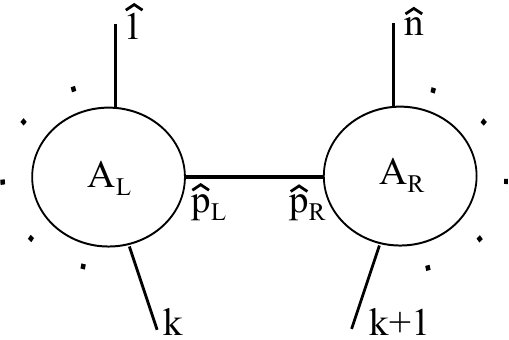}
\caption{The structure of the $k$th BCFW diagram contributing to an $n$ point function.}\label{fig:diagram1}
\end{figure}

Let us elaborate on the sum over intermediate states, which is implicit in the above equation.
To perform this sum it is useful to properly align the little group indices.
For instance, one can take $\hat \lambda_R=i\hat \lambda_L$ and $\hat{\tilde\lambda}_R=-i\hat{\tilde\lambda}_L$,
consistent with $\hat p_R=-\hat p_L$.  Then the bosonic polarisations $\epsilon_{m}$
are the same on each side and the sum over gluon states is simply the contraction of the little group vector indices.
Similarly for the fermion states.
In equations, with careful normalization the contraction is
\begin{equation}
A_{L} A_R \equiv A_{Lm} A_{Rm} - iA_{L\ad}A_{R\ad}
\end{equation}
where the explicit indices are those of the exchanged particle with momentum $\hat p_{L,R}$. Using the supersymmetry transformations
of the amplitudes, for example,
\begin{equation}
\hat Q_L A_{Lm} = \rho_{m \a \ad} A_{L\ad}  \hat \lambda_{L\a}, \quad
\hat Q_L A_{L \ad} = \rho_{m \a \ad} A_{Lm}  \hat \lambda_{L\a},
\end{equation}
it is easy to check that $(\hat Q_L+\hat Q_R) A_L A_R =0$.

\section{Symmetries}
\label{sec:symmetries}

We shall now look for a ten-dimensional symmetry generalizing the dual conformal symmetry
in $\mathcal{N}=4$.

It is useful to first briefly review scale invariance.
Ten-dimensional Yang-Mills is, of course, not scale invariant.
Nevertheless, the classical equations of motion contain no intrinsic scale
and so in a sense the classical theory can be said to be scale-invariant.  The precise sense is that
$n$-point tree amplitudes are homogeneous in a simple way.
This can easily be seen from the fact that the Feynman rules are similar to those in four dimensions.
Thus defining a dilatation generator $d_i=[p_i^\mu \frac{\partial}{\partial p_i^\mu}+1]$ as in four dimensions,
one finds that $n$-point amplitudes are annihilated by the operator
\begin{equation}
  [-4 + \sum_{i=1}^n d_i] A_n = 0. \label{dilatation}
\end{equation}
This is the meaning of ``dilatation symmetry'' in the classical theory; it is simple dimensional analysis.

\subsection{Dual conformal symmetry}

The dual conformal symmetries are defined in the planar limit of the theory.
They are best stated in terms of the region momenta $x_i$ defined by
  $p_i= x_{i}-x_{i{-}1}$; 
the $x_i$ are defined up to an overall shift.
For an infinitesimal vector $v^\mu$ the dual conformal transformation is then
\begin{equation}
  \delta_v x_i^\mu = v{\cdot}x_i x_i^\mu  - \frac12 x_i^2 v^\mu.  \label{defDCI}
\end{equation}

To obtain a symmetry of the theory this must actually be supplemented by
a transformation of fermionic variables.
In ten-dimensional Yang-Mills we will only be able to describe this in operator form,
since we are not using a superspace formulation.
We will follow closely \cite{drummondhennplefka}.
In order to proceed we must choose an explicit representative for the $x_i$: we make the convenient
choice $x_i=\sum_1^i p_i$.
Corresponding to this choice we define the generator:
\begin{equation}
  C^\mu \equiv \sum_{j\leq i} (1-\frac12\delta_{ij}) \left[ p_j^\nu \Lambda^\mu_{i\nu} + p_j^\mu d_i
   + \frac18 \sigma^\mu_{ab} Q_j^a Q_i^b\right].  \label{DCIgen}
\end{equation}

This does not come out of nowhere.  Largely, it is inspired from \cite{drummondhennplefka}.
The bosonic terms are crafted so as to reproduce the transformation (\ref{defDCI}) for the $p_i$:
\begin{equation}
 \delta_v p_i^\mu \equiv \delta_v (x_i^\mu-x_{i{-}1}^\mu) =
 x_i^\mu v{\cdot}p_i + v{\cdot}x_i p_i^\mu - x_i{\cdot}p_i v^\mu - p_i^\mu v{\cdot}p_i. \nonumber
\end{equation}
This is reproduced by the generator (\ref{DCIgen}),
\begin{equation}
 [v{\cdot}C, p_i^\mu] = \sum_{j\leq i} (1-\frac12\delta_{ij})
   \left[ p_j^\mu v{\cdot}p_i - p_j{\cdot}p_i v^\mu + v{\cdot}p_j p_i^\mu\right]
 =
 x_i^\mu v{\cdot}p_i + v{\cdot}x_i p_i^\mu - x_i{\cdot}p_i v^\mu - p_i^\mu v{\cdot}p_i,
 \nonumber
\end{equation}
where we have used $\sum_1^i p_j=x_i$.

The fermionic term is the only dimension-one Lorentz vector one could write down.
Its normalization could be obtained by comparing with the four-dimensional formula
given in \cite{drummondhennplefka}, to which our formula has to reduce
in cases where all particles and the index $\mu$ lie in a four-dimensional subspace, as
we demonstrate in Appendix~\ref{AppendixOnReduction}.
However, it is also possible to check the normalization directly
using either of the following two physical requirements.  

The first of these requirements is that (\ref{DCIgen}) be cyclic invariant:
if a particle other than particle $n$ were chosen as the origin, $x_n=0$, the difference should
be a symmetry.
In general, under a change of origin, sums $\sum_{j\leq i} A_jB_i$ as considered here change
by the amount
% \cite{drummondhennplefka}
\begin{equation}
 \sum_{2\leq j<i\leq n{+}1} A_j B_i - \sum_{j<i} A_j B_i = (\sum_i A_i) B_1 - A_1 (\sum_i B_i). \nonumber
\end{equation}
(The contribution from $i=j$ does not contribute to the change.)
Here, $A$ and $B$ are various combinations of $p,Q$ and $\Lambda,d,Q$, respectively.
Since we do not include momentum conservation $\delta$-functions in our amplitudes,
$\sum_i p_i =0$ when appearing on the left of all derivative operators.
Thus the first term vanishes except for its fermionic component.
Similarly, $\sum_i \Lambda_i \simeq \sum_i Q_i \simeq 0$ and $\sum_i d_i \simeq 4$ when
acting on the right, according to (\ref{dilatation}).
This reduces the second term to $-4p_1^\mu$.
We thus find the change in $C^\mu$, modulo symmetries:
\begin{equation}
 \frac18  \sigma^\mu_{ab} \{ Q_1^a, Q_1^b \} -4p_1^\mu = 0.  \label{cancel14}
\end{equation}
The first term is equal to $\frac18 \mbox{Tr}~ [2\sigma^\mu
/\!\!\!p_1]=4p_1^\mu$, and so one finds that $C^\mu$ is indeed cyclically
invariant. As mentioned above, this computation can be used to fix
the normalization of the fermionic term in (\ref{DCIgen}). 
% Notice that
%these algebraic manipulations are unchanged in any dimension $D\leq 10$.

A second important property of $C^\mu$,
is that it commutes with supersymmetry: $[C^\mu,Q^a]= 0$.
Proof of that statement will be omitted.  Again, this holds only if the coefficient of the fermionic term
is as in (\ref{DCIgen}).
% More precisely, $[C^\mu,q^A] = \frac14 (/\!\!\!p \gamma^mu q)^A$.

In summary, we have defined the transformation (\ref{DCIgen}).  It has a sensible bosonic part,
it is cyclic invariant, and commutes with supersymmetry.  As such it stands a good chance at being a symmetry of the theory.
Furthermore, by construction it reduces to the four-dimensional generator
in \cite{drummondhennplefka}, when all momenta are the $\mu$ index lie in a four-dimensional subspace.
It follows that the three-point function is invariant\footnote{To make this argument completely precise we may assume that all momenta are
 in a null-plane in four-dimensions spanned by two orthogonal null vectors $p_1$ and $p_2$.  From four-dimensional physics,
 this three-point vertex is annihilated by $v\cdot C$ for $v^\mu$ in four dimensions. By ten-dimensional rotational invariance it follows
 that $v\cdot C A_3=0$ also for any linear combination of vectors obtained from rotations leaving $p_1$ and $p_2$ unchanged;
 such vectors span the whole ten-dimensional space, leading to $v\cdot C A_3=0$ for any $v$.}.

\subsection{Symmetry of BCFW terms}

To confirm that our generator is indeed a symmetry of the theory, we use the BCFW construction
of the tree S-matrix (\ref{treeBCFW}).  We will see
that each term manifestly preserves the symmetry.  This is similar
to what was found in \cite{brandhuber} in the context of $\mathcal{N}=4$.

We continue using the BCFW deformation along particles $1$ and $n$
\begin{equation}
 \hat\lambda_{n\a} = \lambda_{n\a} + zM_{\a\b} \lambda_{1\b}, \quad
 \hat\lambda_{1\b} = \lambda_{1\b} - zM_{\a\b} \lambda_{n\a}, \nonumber
\end{equation}
so that $\hat p_1= p_1-zq$.
The general term in the BCFW formula,
if $P=\sum_1^k p_i$ is the exchanged momentum,  is then of the form
\begin{equation}
  A_{(k)} = \frac{e^{-\frac{P^2}{4P{\cdot}q} \Gamma_n M \Gamma_1}}{P^2}
    A_L(\hat p_1,\ldots,p_k,\hat p_{k+\frac12^-})
    A_R(\hat p_{k+\frac12^+},p_{k{+}1},\ldots,\hat p_n)  \label{BCFWtree2}
\end{equation}
where $\hat p_1\equiv p_1 {-} \frac{P^2}{2P{\cdot}q} q$ and
$
 \hat p_{k{+}\frac12^\pm} \equiv \pm(P {-} \frac{P^2}{2P{\cdot}q} q)$.
This notation, ``$\hat p_{k{+}\frac12^\pm}$'', is slightly awkward but it will prove convenient in what follows.

\subsubsection{Bosonic terms}

We want to rewrite our generator in a way that makes its action
on the factors $A_L$ and $A_R$ manifest.
Our first step is to do that for the bosonic terms.

We define our shift by picking a particular $M$; then the deformation vector $q$ depends implicitly on $\lambda_1$ and $\lambda_n$.
%(we view $M$ as not depending on $\lambda$, which is permitted)
Therefore, it is useful to begin by expressing $C^\mu$ in terms of these variables.
For the bosonic terms this gives
\begin{equation}
 C^\mu_\textrm{bosonic} = \sum_{j\leq i} [1-\frac12 \delta_{ij}] \left[
 \frac{ [/\!\!\!p_j \sigma^\mu   \lambda_{i\alpha}]^a}{2}
 \frac{\partial}{\partial \lambda_{i\alpha}^a}
+ (\lambda_i\to \tilde\lambda_i) + p_j^\mu \right]. \nonumber
\end{equation}

We now use the chain rule to write the terms containing derivatives in terms of the variables entering $A_L$ and $A_R$.
One can easily compute that $[C^\mu, q^\nu] = \frac{p_1^\mu-p_n^\mu}{2} q^\nu$
and that $[C^\mu, P^\nu]= P^\mu P^\nu - \frac12\delta^{\mu\nu} P^2$.
Using these results, one can further compute that
\begin{eqnarray}
 [C^\mu, \hat \lambda_{1\a}]
  &=& 
  \frac12 \hat p_1^\mu \hat\lambda_{1\a}
  + \frac{P^2}{2P{\cdot}q} \frac{ (\hat\lambda_{n\gamma} M_{\gamma[\a} \sigma^\mu \hat\lambda_{1\b]}) \hat \lambda_{1\b}}{4}
  + \frac{P^2}{2P{\cdot}q} \frac{ /\!\!\!q \sigma^\mu \hat \lambda_{1\a}}{2},
 \nl
 ~[C^\mu, \lambda_{i\a}] &=&
\sum_{j\leq i} [1-\frac12\delta_{ij}]
\frac{ /\!\!\!\hat p_j \sigma^\mu \lambda_{i\a}}{2}
  + \frac{P^2}{2P{\cdot}q} \frac{ /\!\!\!q \sigma^\mu \lambda_{i\a}}{2},
  \nl
 ~[C^\mu, \hat \lambda_{k{+}\frac12^\pm \a}] &\equiv&
  \pm \frac12 \hat p_{k{+}\frac12^\pm}^\mu   \hat \lambda_{k{+}\frac12^\pm \a}^A
  + \frac{P^2}{2P{\cdot}q} \frac{ /\!\!\!q \sigma^\mu \hat \lambda_{k{+}\frac12^\pm \a}}{2},
\nl
 ~[C^\mu, \hat \lambda_{n\a}]
  &=&
  -\frac12 \hat p_n^\mu \hat\lambda_{n\a}
  + \frac{P^2}{2P{\cdot}q} \frac{ (\hat\lambda_{n[\b} M_{\a]\gamma} \sigma^\mu \hat\lambda_{1\gamma}) \hat\lambda_{n\b}}{4}
  + \frac{P^2}{2P{\cdot}q} \frac{ /\!\!\!q \sigma^\mu \hat \lambda_n}{2}.
 \label{arrayofjunk}
\end{eqnarray}
The third of these equations requires some comment. Since $\hat p_{k + \frac12^\pm} = \pm ( P - \frac{P^2}{2 P \cdot q q})$,
all we can require is that the action of $C^\mu$ on $\hat \lambda_{k + \frac12^\pm}$ be consistent with the action
of $C^\mu$ on $ \pm ( P - \frac{P^2}{2 P \cdot q q})$.  It is also useful to impose the
constraint $\delta \hat \lambda_{k + \frac12^+} = i \delta \hat \lambda_{k + \frac12^-}$ as was done above.
The transformations of $\hat \lambda_{k + \frac12^\pm}$
are then determined up to a common and irrelevant little group rotation; the choice in (\ref{arrayofjunk})
fulfills all these conditions.  Let us also add that the orthogonality relations (\ref{orthoQ})
are quite useful in proving these equations.
Their physical interpretation will be given shortly.

These relations allow us to write the terms containing derivatives as:
\begin{align}
  \sum_{j\leq i} [1-\frac12 \delta_{ij}]
\frac{ [/\!\!\!p_j  \sigma^\mu   \lambda_{i\alpha}]^a}{2}
& \frac{\partial}{\partial \lambda_{i\alpha}^a}
  =
\left[ \sum_{1\leq j\leq i \leq k{+}\frac12^-}
+ \sum_{k{+}\frac12^+ \leq j\leq i \leq n}\right]
 [1-\frac12\delta_{ij}]
 \frac{ [/\!\!\!\hat p_j \sigma^\mu  \hat \lambda_{i\alpha}]^a}{2}
  \frac{\partial}{\partial \hat \lambda_{i\alpha}^a} 
 \nonumber \\
&+
\frac{P^2}{2P{\cdot}q} \frac{ (\hat\lambda_{n\gamma} M_{\gamma[\a} \sigma^\mu \hat\lambda_{1\b]}) \hat \lambda_{1\b}^a}{4}
    \frac{\partial}{\partial \hat \lambda_{1\a}^a}
 + \frac{P^2}{2P{\cdot}q} \frac{ (\hat\lambda_{n[\b} M_{\a]\gamma} \sigma^\mu \hat\lambda_{1\gamma}) \hat\lambda_{n\b}^a}{4}
    \frac{\partial}{\partial \hat \lambda_{n\a}^a}
\nonumber    \\
&+
\frac{P^2}{2P{\cdot}q} \sum_i \frac{ [/\!\!\!q \sigma^\mu \hat \lambda_{i\alpha}]^a}{2}
  \frac{\partial}{\partial \hat \lambda_{i\alpha}^a}.  \label{bigbosonic}
\end{align}
plus an identical equation with antichiral spinors.  

This expression looks complicated but actually it is of the form we are looking for.
All terms arise from the different terms in (\ref{arrayofjunk}), and each has a transparent physical interpretation.
Indeed, the first line contains just the naive transformation 
law for each of the sub-amplitudes $A_L$ and $A_R$.  The second line is a set of little-group rotations,
which is devoid of any effect on physical kinematics.
The last line is interpreted as an overall shift in the $x$ space origin
from $x_n=0$ to $x_n=\frac{P^2}{2P{\cdot}q} q$, which again has no effect on the physical kinematics.
It is thus permitted, at this stage, to anticipate that the last two lines will combine with other
fermionic terms in order to produce zero.

To finish computing the bosonic term we only have to include the inhomogeneous term.
\begin{equation}
  \sum_{j\leq i}[1-\frac12\delta_{ij}]  p_j^\mu \to
\left[ \sum_{1\leq j\leq i \leq k{+}\frac12^-}
+ \sum_{k{+}\frac12^+ \leq j\leq i \leq n}\right]
  [1-\frac12\delta_{ij}]  \hat p_j^\mu - 2\hat p_{k{+}\frac12^+}^\mu + (n{-}2) \frac{P^2q^\mu}{2P{\cdot}q}. 
 \label{bosonicsmall}
\end{equation}
The arrow means that we have commuted the whole operator $C^\mu$ across the factor $1/P^2$,
which has added $-P^\mu$ to the right.
Using (\ref{dilatation}) and $\sum_i \hat\Lambda_i\simeq 0$,
the last line in (\ref{bigbosonic}) and the last term in (\ref{bosonicsmall}) are shown to sum up to
$+4 \frac{P^2q^\mu}{2P{\cdot}q}$.

\subsubsection{Fermionic terms}

The next step is to compute the fermion terms.  It is useful to introduce hatted gamma matrices $\hat \Gamma_1$, $\hat \Gamma_n$
\begin{equation}
\hat \Gamma_1 = \Gamma_1 - \frac{P^2}{2P{\cdot}q} M\Gamma_n,\quad
\hat \Gamma_n = \Gamma_n + \frac{P^2}{2P{\cdot}q} M\Gamma_1,
\end{equation}
so that
\begin{equation}
    Q_1^a
  = \hat Q_1^a + \frac{P^2}{2P{\cdot}q} ( 
  \hat\lambda_1^a M \hat\Gamma_n 
+ \hat\lambda_n^a M \hat\Gamma_1), \quad
    Q_n^a
  = \hat Q_n^a - \frac{P^2}{2P{\cdot}q} ( 
  \hat\lambda_1^a M \hat\Gamma_n 
+ \hat\lambda_n^a M \hat\Gamma_1). \quad
  \nonumber
\end{equation}
(Hatted operators are defined by $\hat Q_1\equiv \hat{\lambda}_1\hat \Gamma_1$.)
The hatted operators are those which act nicely on the subamplitudes $A_L$ and $A_R$ after they are
commuted across the fermion exponential.
Using this substitution to express the fermion terms in terms of hatted variables one finds
\begin{eqnarray}
\label{fermionAction}
  \sum_{j\leq i} [1-\frac12 \delta_{ij}]  \frac18  Q_j \sigma^\mu Q_i
  &= &
\left[ \sum_{1\leq j\leq i \leq k{+}\frac12^-}
+ \sum_{k{+}\frac12^+ \leq j\leq i \leq n}\right]
 [1-\frac12\delta_{ij}]  \frac18  \hat Q_j \sigma^\mu \hat Q_i
 \nl
 &+&
\frac18
\left[\hat Q_{k{+}\frac12^+} - \sum_{1\leq j\leq k} \hat Q_j \right] \sigma^\mu
\left[\hat Q_{k{+}\frac12^-} - \sum_{k{+}1\leq i\leq n} \hat Q_i\right] - \frac14
\hat Q_{k{+}\frac12^+}\sigma^\mu\hat Q_{k{+}\frac12^-}
\nl
&-& 
\frac{P^2}{2P{\cdot}q}  \frac{\hat\lambda_{n\gamma} M_{\gamma[\a} \sigma^\mu \hat\lambda_{1\b]}}{16} \hat\Gamma_1^\a \hat\Gamma_1^\b
-
\frac{P^2}{2P{\cdot}q}  \frac{\hat\lambda_{n[\b} M_{\a]\gamma} \sigma^\mu \hat\lambda_{1\gamma}}{16} \hat\Gamma_n^\a \hat\Gamma_n^\b
\nl
&+& 2\hat p_{k{+}\frac12^+}^\mu - 4 \frac{P^2q^\mu}{2P{\cdot}q}.
%-\frac{P^2}{2P{\cdot}q} 4q^\mu + (2P^\mu -\frac{P^2}{2P{\cdot}q} 2q^\mu)
\end{eqnarray}
The hatted operators have the property that $\hat Q_{k{+}\frac12^-} + \hat Q_{k{+}\frac12^+}=0$ when acting on the left,
and that $\hat Q_{k{+}\frac12^-} - \sum_{k{+}1\leq i\leq n} \hat Q_i=0$ when acting on the right.
Using this property one can show that the second line vanishes on amplitudes.
The third line is a set of little-group rotations
which nicely complete the rotations present in (\ref{bigbosonic}).  In order to find this form, without cross-terms $\hat\Gamma_1\hat\Gamma_n$,
it is important to include also terms which arise from the bosonic derivatives in $C^\mu$ acting on the exponential.
Finally, the fourth line collects various anti-commutators which arose along the way.
We observe that it perfectly cancels against the inhomogeneous terms in (\ref{bosonicsmall}).

\subsubsection{Conclusion of the proof}

Adding up the results of the preceding two subsections, what we have found is that
$C^\mu$ acting on a BCFW term $A_{(k)}$ is equal to
\begin{equation}
 C^\mu[p_1,\ldots,p_n] A_{(k)} =
 \frac{e^{-\frac{P^2}{4P{\cdot}q} \Gamma_n M \Gamma_1}}{P^2}
\left(  C^\mu[\hat p_1,\ldots,\hat p_{k{+}\frac12^-}]
  +C^\mu[\hat p_{k{+}\frac12^+},\ldots,\hat p_n]
 +X^\mu\right) A_L A_R,
\nonumber
\end{equation}
where\footnote{We discuss $X$ more in Appendix~\ref{appendixOnX}.}
\begin{equation}
 X^\mu =
 \frac{P^2}{2P{\cdot}q}  \frac{\hat\lambda_{n\gamma} M_{\gamma[\a} \sigma^\mu \hat\lambda_{1\b]}}{8} \hat S_{1\a\b}
 + \frac{P^2}{2P{\cdot}q}  \frac{\hat\lambda_{n[\b} M_{\a]\gamma} \sigma^\mu \hat\lambda_{1\gamma}}{8} \hat S_{n\a\b}.
\nonumber
\end{equation}
The first two terms in the bracket are dual conformal generators acting on
$A_L$ and $A_R$.  By the induction hypothesis, they vanish.
The third term, $X^\mu$, is just the ten dimensional analogue of the usual four dimensional helicity operator, and
thus annihilates the amplitudes on its own.
This means that we have shown that
\begin{equation}
 C^\mu[p_1,\ldots,p_n] A_{(k)} = 0
\end{equation}
for each individual BCFW term,
completing our proof of the dual conformal invariance of the tree S-matrix.

\section{Discussion and Outlook}
\label{sec:concl}

In this paper we have introduced a ten-dimensional spinor-helicity formalism which can be used to describe the scattering
amplitudes of massless particles with spin.  The formalism is compatible with supersymmetry in a simple way,
and we have discussed supersymmetric BCFW recursion relations for tree amplitudes in $\mathcal{N}=1$ super Yang-Mills.
A remarkable feature of these recursion relations is that each term manifests a symmetry of the theory,
hidden from the Lagrangian formulation, and similar to the dual conformal symmetry known in four-dimensional $\mathcal{N}=4$ super Yang-Mills.

The reader may be left with the impression that there is some magic going on in the computations of the preceding section.
From our viewpoint, the real magic is not to be found here but
in the four-dimensional computations \cite{Drummond:2008vq,drummondhennplefka,brandhuber}.
Let us explain what we mean by that.

We have defined a dual conformal generator, $C^\mu$, (\ref{DCIgen}), which acts on ten-dimensional super-amplitudes.
It can be restricted to amplitudes ``living'' in a four-dimensional subspace,  in which case,
when the index $\mu$ is restricted to the same subspace, it reduces
to the four-dimensional generator in \cite{drummondhennplefka}.
That generator is known to annihilate BCFW terms.  Thus, even before we began,
massive algebraic cancellations were guaranteed to occur in our computation,
so as to reproduce the four-dimensional results.
The purpose of our computations was to make sure that these cancellations did not depend on a four-dimensional
subspace being special.  In some sense,
given that all our formulas are ten-dimensionally covariant, this doesn't look, with hindsight, too surprising.

We would like to conclude this paper with some open questions and open speculations.
\begin{itemize}
\item Pragmatically speaking, the enhanced symmetry means that ten-dimensional amplitudes
 should be simpler than expected.  This should help in writing them down,
 if only one knew a formulation in which the symmetry generators are simple.

\item In four dimensions, a geometric realization of the symmetry is provided by
   a duality between the scattering amplitudes and certain polygonal Wilson loops \cite{Alday:2007hr,Drummond:2007aua,Brandhuber:2007yx,Berkovits:2008ic}.  Recently this duality was extended to super-amplitudes
   \cite{Mason:2010yk,CaronHuot:2010ek,Brandhuber:2010mi}.
  One can ask whether some analog of this duality might exist in other dimensions; that would provide a geometric meaning to the symmetry.
  
  A priori such a duality would appear to be outrageous.  For one thing, super Yang-Mills theory in higher dimensions is not conformal,
  so how would such a duality explain the conformal symmetry?  However, in the four-dimensional case, it turns out that
  a proper subsector of the theory (the self-dual sector) suffices to compute the Wilson loop to the accuracy corresponding to tree amplitudes
  \cite{Mason:2010yk,CaronHuot:2010ek}.
  There this is related to a massive symmetry enhancement taking place at tree level (to a Yangian). Perhaps a similar miracle could occur
  in higher dimensions.
  
  Another issue is that the Yang-Mills propagators in higher-dimensions are too singular for a Wilson loop to match with
  a scattering amplitude. However, all this shows is that the dual of a scattering amplitude might be some brane.  Perhaps also the dual theory is not super Yang-Mills.
  Although we have no concrete proposal to make, we cannot single-handedly rule out such a duality.  

\item We have defined the generator only for color-ordered partial amplitudes.  For a $n$ point amplitude the definition of such amplitudes
 requires that $N_c\geq n$, so in a sense we are working in the large-$N_c$ limit.  This is puzzling.
  Indeed, the large-$N_c$ limit is usually invoked to render the dynamics more classical,
  but since we are considering classical dynamics anyway it is not clear what this limit is buying us here.
  So, perhaps the symmetry generator could in addition be defined algebraically at finite $N_c$.
%  Technically, the planar limit is useful to define the $x_i$ coordinates which give the symmetry its geometrical picture,
%  but whether it is needed to define the symmetry algebraically is not so clear.
  We would thus like to conclude with a small challenge to the reader: either to extend the generator to arbitrary gauge group,
  or to conclusively show that this is impossible.
\end{itemize}

\acknowledgements

SCH gratefully acknowledges support from NSF grant PHY-0969448. DOC gratefully acknowledges support from a Martin and Helen A. Chooljian membership at the IAS, and by DOE grant DE-FG02-90ER-40542.

\appendix

\section{Symmetry algebra}

It is useful to give the symmetry algebra of the theory, if only to fix our conventions. The supersymmetry algebra is
\begin{align}
 [\Lambda^\mu{}_\nu, p^\sigma] &= p^\mu \delta_\nu^\sigma - p_\nu \eta^{\mu\sigma}, &
 [\Lambda^\mu{}_\nu, Q^a] &= -\frac14 \left( \sigma^\mu \tilde \sigma^\nu - \sigma^\nu \t \sigma^\mu \right)^a_b ~Q^b 
\\
 \{Q^a,Q^b\}&= 2\tilde \sigma^{ab}_\mu p^\mu,&  [p^\mu, Q^a] &= 0.
\end{align}
The commutator of $\Lambda$ with any other object with Lorentz indices, including itself,
is as usual, and other commutator vanishes.
The generator $C^\mu$ obeys in addition the commutation relation
\begin{align}
 [\Lambda^\mu{}_\nu, C^\sigma] &= C^\mu \delta_\nu^\sigma - C_\nu \eta^{\mu\sigma}, &
 [C^\mu,p^\nu]&= [C^\mu,Q^a]=[C^\mu,C^\nu]=0,
\end{align}
and the dilatation generator $d$ obeys $[d,p^\mu]=p^\mu$, $[d,Q^a]=\frac12 Q^a$ and $[d,C^\mu]=C^\mu$.

We notice, in particular, that these generators form a closed algebra; thus, the symmetry algebra of the theory does not
extend to an infinite-dimensional one as was the case in four dimensions.

The dual conformal generator $C^\mu$ can be naturally thought of as part of a SO(2,10) conformal group.  The extension is somewhat trivial, actually, since the $x$-space translations do not actually act on amplitudes.  They do not do anything.
By combining these translations, $C^\mu$, the dilatation generator, and Lorentz transformations one obtains a (``dual'') SO(2,10) in the familiar way,
The definition of the dilatation and dual conformal generators acting on an $n$ point amplitude are
\begin{align}
d &= \sum_{i=1}^n [p_i^\mu \frac{\partial}{\partial p_i^\mu}+1] \\
C^\mu &= \sum_{j\leq i} (1-\frac12\delta_{ij}) \left[ p_j^\nu \Lambda^\mu_{i\nu} + p_j^\mu d_i
   + \frac18 \sigma^\mu_{ab} Q_j^a Q_i^b\right].  
\end{align}
We also find it convenient define the generator of little group rotations
\begin{equation}
 S_{\a\b} = 
 \lambda_{[\a}^a \frac{\partial}{\partial \lambda_{\b]}^a}
 +\frac12 \rho_{\a \b}{}^{\ad \bd} \tilde\lambda_{[\ad a} \frac{\partial}{\partial \tilde \lambda_{\bd] a}}
 + \frac14 [\Gamma_\a, \Gamma_\b].
\end{equation}
where
\begin{equation}
\rho_{\a\b}{}^{\ad\bd} = \frac 14 (\rho_{m \a \ad} \rho_{m \b \bd} - \rho_{m \b \ad} \rho_{m \a \bd})
\end{equation}
relates the adjoint representations obtained from the various spinor representations.  

\section{Appendix on $S$}
\label{appendixOnX}

In this Appendix we would like to check some properties of the little-group rotations $S$, in particular with respect to
the way it arises in the BCFW computation. Indeed the little group rotations (going into $X^\mu$) which arise in
Eq.s~\eqref{bigbosonic} and \eqref{fermionAction} are of the form
\begin{multline}
\frac{ (\hat\lambda_{n\gamma} M_{\gamma[\a} \sigma^\mu \hat\lambda_{1\b]}) \lambda_{1\a}^a }{4}
    \frac{\partial}{\partial \hat \lambda_{1\b}^a}
+    \frac{ (\hat{\tilde \lambda}_{n\cd} \tilde M_{\cd[\ad} \tilde \sigma^\mu \hat{\tilde \lambda}_{1\bd]})
   \hat{\tilde \lambda}_{1\ad a}}{4} \frac{\partial}{\partial \hat{\tilde \lambda}_{1\bd a}}
+    \frac{\hat\lambda_{n\gamma} M_{\gamma[\a} \sigma^\mu \hat\lambda_{1\b]}}{16} \hat\Gamma_1^\a \hat
     \Gamma_1^\b.
\end{multline}
This will be equal to the form claimed in the main text, that is, $x_{\a\b}^\mu S_{\a\b}$ where
\begin{equation}
x_{\a \b}^\mu \equiv  \frac{ \hat\lambda_{n\gamma} M_{\gamma[\a} \sigma^\mu \hat\lambda_{1\b]}}{8},
\end{equation}
provided it is true that
\begin{equation}
x_{\ad \bd}^\mu \equiv  \frac{ \hat{\tilde\lambda}_{n\cd} \tilde M_{\cd[\ad} \tilde\sigma^\mu \hat{\tilde\lambda}_{1\bd]}}{8}
\end{equation}
is appropriately related to $x^\mu_{\a\b}$ through $\rho_{\a\b}{}^{\ad\bd}$.

The basic reason why this is true is that there is only one adjoint representation. We may label adjoint objects by antisymmetric indices of any of the three dimensional representations; then there are change of basis matrices relating these labellings. The objects $\rho_{\a\b}{}^{\ad\bd}$ can be thought of as such a change of basis matrix. 

Let us begin by simplifying the object $x$ a little. In particular, it is straightforward to see that we may write $x$ as
\begin{equation}
x^\mu_{\a \b} = \frac{ \lambda_{n\gamma} M_{\gamma[\a} \sigma^\mu \lambda_{1\b]}}{8}
= \frac18 \epsilon_m^\mu \rho_{m \b \ad} (\tilde \lambda_{1 \ad} \cdot \lambda_{n \c}) M_{\c \a} - (\alpha \leftrightarrow \beta),
\end{equation}
where the polarisation vector is in gauge $p_n$. Now, as we observed in the text, the object $V_{\a \ad} = (\tilde \lambda_{1 \ad} \cdot \lambda_{n \b}) M$ does not contain any little group 3-form. Thus, $V_{\a \ad} = \rho_{m \a \ad} V_m$. Consequently,
\begin{equation}
x_{\a \b}^\mu = \frac18 (\rho_{m \b \ad} \rho_{n \a \ad} - \rho_{m \a \ad} \rho_{n \b \ad}) \epsilon^\mu_m V_n = 
-\frac{1}{2} \rho^{mn}{}_{\a \b} \epsilon^\mu_m V_n.
\end{equation}
Similarly, we find
\begin{equation}
x_{\ad \bd}^\mu = -\frac{1}{2} \rho^{mn}{}_{\ad \bd} \epsilon^\mu_m V_n.
\end{equation}
Of course, it must be true that $\rho^{mn}{}_{\ad \bd}$ and $\rho^{mn}{}_{\a \b}$ are proportional; we find that
\begin{equation}
\rho^{mn}{}_{\ad \bd} = \frac12 \rho^{mn}{}_{\a \b} \rho^{\a \b}{}_{\ad \bd}
\end{equation}
which completes the proof. 

\section{Reduction to four dimensions}
\label{AppendixOnReduction}

Let us reduce the generator $C^\mu$ to four dimensions. We take $\mu=0{,}1{,}2{,}3$ and
assume that all momenta are in that subspace. Let us first recall our original, ten-dimensional generator
\begin{align}
C^\mu&= \sum_{j\leq i} (1-\frac12\delta_{ij}) \left[ p_j^\nu \Lambda^\mu_{i\nu} + p_j^\mu d_i
   + \frac18 \sigma^\mu_{ab} Q_j^a Q_i^b\right]. 
\nonumber \\
&= \sum_{j\leq i} (1-\frac12\delta_{ij}) \left[  \frac{ [/\!\!\!p_j \sigma^\mu   \lambda_{i\alpha}]^a}{2}
 \frac{\partial}{\partial \lambda_{i\alpha}^a}
+ (\lambda_i\to \tilde\lambda_i) + p_j^\mu
   + \frac18 \sigma^\mu_{ab} Q_j^a Q_i^b\right].  \label{10dform}
\end{align}
Upon restricting to four dimensions $C^\mu\to C^{\alpha\dot\alpha}$, following the notations in \cite{drummondhennplefka},
this becomes
\begin{equation}
C^{\alpha\dot\alpha} =
\sum_{j\leq i} (1-\frac12\delta_{ij}) \left[
\tilde\lambda_j^{\ad} \lambda_i^\alpha \lambda_j^\beta \frac{\partial}{\partial \lambda_i^\beta}
+\lambda_j^{\a} \tilde \lambda_i^{\ad} \tilde \lambda_j^\bd \frac{\partial}{\partial \tilde \lambda_i^\bd}
 + \lambda_j^\a \tilde\lambda_j^\ad
 + \frac12 \lambda_j^\a \tilde \lambda_i^\ad \eta_j^A \frac{\partial}{\partial \eta_i^A}
 + \frac12 \tilde\lambda_j^\ad \lambda_i^\a \frac{\partial}{\partial \eta_j^A} \eta_i^A\right].  \label{4dform}
\end{equation}
Some comments are in order. The first three terms and their relative normalization are relatively straightforward and unmistakable.
The form of the fermion term is also rather unmistakable, because particle $j$ always acts on the left of particle $i$
and the original ten-dimensional formula is non-chiral. The index $A=1\ldots4$ is the SU(4)${}_R$ R-symmetry index.
Thus this is the only possible from. One might inquire as to its normalization.
One foolproof check is to take $i=j$ in (\ref{10dform}), which reduces the fermion term to $2p_i^\mu$.
Since the same is true in (\ref{4dform}) (with $2p_i^\mu \leftrightarrow 2\lambda_i^\a\tilde\lambda_i^\ad$)
we conclude that all terms are properly normalized.

Thus (\ref{4dform}) is the dimensional reduction of $C^\mu$. It is very close to, but not quite equal to, equation (22) in \cite{drummondhennplefka}. Let us now show that it is equivalent to it. The first step is to add to it the product of symmetries
\begin{equation}
 -\frac12 \sum_{i,j} \tilde\lambda_j^\ad \lambda_i^\a \frac{\partial}{\partial \eta_j^A} \eta_i^A
\end{equation}
which obviously annihilates scattering amplitudes. Our generator then becomes
\begin{align}
C^{\alpha\dot\alpha} &=
\sum_{j<i} \left[
\tilde\lambda_j^{\ad} \lambda_i^\alpha \lambda_j^\b  \frac{\partial}{\partial \lambda_i^\b}
+\lambda_j^{\a} \tilde \lambda_i^{\ad} \tilde \lambda_j^\bd \frac{\partial}{\partial \tilde \lambda_i^\bd}
 + \lambda_j^\a \tilde\lambda_j^\ad
 + \lambda_j^\a \tilde \lambda_i^\ad \eta_j^A \frac{\partial}{\partial \eta_i^A}\right] \nonumber \\
 &\quad +  \sum_i \frac12\lambda_i^\a \tilde\lambda_i^\ad \left[ 
 \lambda_i^\b \frac{\partial}{\partial \lambda_i^\b}
 +\tilde \lambda_i^\bd \frac{\partial}{\partial \tilde \lambda_i^\bd}
 + 1
  + \frac12 (\eta_i^A \frac{\partial}{\partial \eta_i^A}
  - \frac{\partial}{\partial \eta_i^A} \eta_i^A)
   \right].
\end{align}
The second step is to notice that the last line, using that the helicity operator
$(  \lambda_i^\b\frac{\partial}{\partial \lambda_i^\b}-
 \tilde \lambda_i^\bd\frac{\partial}{\partial \tilde \lambda_i^\bd}
  -\eta_i^A \frac{\partial}{\partial \eta_i^A}
   +2)$ annihilates amplitudes, and that $\sum_i \lambda_i^\a\tilde\lambda_i^\ad=0$, is equal to
\begin{equation}
 \sum_i \lambda_i^\a \tilde\lambda_i^\ad \left[ 
 \tilde \lambda_i^\bd \frac{\partial}{\partial \tilde \lambda_i^\bd}
  + \eta_i^A \frac{\partial}{\partial \eta_i^A} \right].
\end{equation}
This way one finds precise agreement with Equation (22) of \cite{drummondhennplefka}.

Actually, to finish the matching, one must recall that in \cite{drummondhennplefka} amplitudes carry explicit
factors of $\delta^4(\sum_i p_i)$ while our amplitudes do not. (We preferred to strip off these factors
in the present paper, because this simplifies the relation between amplitudes in different dimensions.)
However, one trivially check that this factor
commutes with $C$, and so nothing is lost by stripping it off like we do.

\end{document}